\documentclass[conference]{IEEEtran}

\usepackage[utf8]{inputenc}
\usepackage{cite}
\usepackage{graphicx}
\usepackage{graphics}
\usepackage[cmex10]{amsmath}
\usepackage{amsfonts}
\usepackage{amssymb}
\usepackage[utf8]{inputenc}
\usepackage{booktabs}
\usepackage{array}
\usepackage{soul}
\usepackage{comment}
\usepackage[hyphens]{url}
\usepackage[a4paper, margin=0.7in]{geometry}
\graphicspath{{figs/}}
\usepackage[usenames,dvipsnames]{xcolor}
\usepackage{siunitx}
\usepackage{multirow}
\usepackage[caption=false]{subfig}
\usepackage{url}
\usepackage{color,soul}
\usepackage[utf8]{inputenc}
\usepackage{lipsum}
\usepackage{mathtools, cuted}
\usepackage[export]{adjustbox}
\usepackage{dblfloatfix}    %
\usepackage[english]{babel}
\usepackage{amsthm}
\usepackage{mathbbol}
\usepackage{bbm}
\usepackage[english]{babel}
\usepackage{blindtext}
\usepackage{bm}
\usepackage{cite}
\usepackage{gensymb} %

\DeclareCaptionFont{mysize}{\fontsize{9pt}{12pt}\selectfont} %
\newtheorem{theorem}{Theorem}

\newtheorem{proposition}{Proposition} %
\newtheorem{lemma}{Lemma}%
\newtheorem{assumption}{Assumption}

\def\BibTeX{{\rm B\kern-.05em{\sc i\kern-.025em b}\kern-.08em
    T\kern-.1667em\lower.7ex\hbox{E}\kern-.125emX}}

\definecolor{bblue}{rgb}{0.67, 0.9, 0.93}
\title{Visual Tool for Assessing Stability of DER Configurations on Three-Phase Radial Networks\\} %

\author{\IEEEauthorblockN{Jaimie Swartz* \quad Brittany Wais \quad Alexandra von Meier}
\IEEEauthorblockA{\textit{Department of Electrical Engineering and Computer Science} \\
\textit{University of California - Berkeley}\\
Berkeley, California, USA \\
*jaimie.swartz@berkeley.edu}
\and
\IEEEauthorblockN{Elizabeth Ratnam}
\IEEEauthorblockA{\textit{College of Engineering and Computer Science} \\
\textit{Australian National University}\\
Canberra, Australia \\
}}

\begin{document}
\maketitle

\begin{abstract}
We present a method and tool for evaluating the placement of Distributed Energy Resources (DER) on distribution circuits in order to control voltages and power flows. 
Our previous work described Phasor-Based Control (PBC), a novel control framework where DERs inject real and reactive power to track voltage magnitude and phase angle targets. Here, we employ linearized power flow equations and integral controllers to develop a linear state space model for PBC acting on a three-phase unbalanced network. We use this model to evaluate whether a given inverter-based DER configuration admits a stable set of controller gains, which cannot be done by analyzing controllability nor by using the Lyapunov equation. Instead, we sample over a parameter space to identify a stable set of controller gains. %
 Our stability analysis requires only a line impedance model and does not entail simulating the system or solving an optimization problem. We incorporate this assessment into a publicly available visualization tool and demonstrate three processes for evaluating many control configurations on the IEEE 123-node test feeder (123NF). 

\end{abstract}

\section{Introduction}

On distribution grids it is increasingly important to effectively control DERs, especially in order to address power quality issues introduced by uncontrolled solar resources. As more controllable DERs are installed, we must analyze whether these devices will co-operate as intended, or if they will exhibit adverse interactions. This coordination problem is not only determined by each DER's control strategy, but also their placement and communication setup. 

Typically, the communication setup for inverter control is local, where inverters use measurements at their own node to modify their power output with the goal of regulating voltage and/or frequency. Yet simulations of local droop volt-var control (DVVC) inverters have indicated risks of oscillations \cite{sudipta_hunting}, reminding us that when coordinating groups of DERs it is critical to guarantee voltage stability. Additionally, there are opportunities for groups of electrically spaced inverters to collectively address voltage issues at nearby nodes.  
PBC is a versatile framework for recruiting diverse DERs to support safe and resilient grid operations. Our team has previously presented the conceptual rationale for PBC \cite{pbc_journal} and demonstrated it with simulations \cite{GA_paper} and with hardware \cite{Gabe_DEGC}. In this work, we develop a state space model under the PBC control framework. We then incorporate this model into a tool that illustrates good locations to place controllable inverter-based DERs. The tool's state space model can be adapted to other power injection setups, such as DVVC and volt-watt control.

Our tool is distinct from capacity maps, which are used to determine when uncontrollable solar PV violates voltage constraints with and without controllable DERs. In the hosting capacity analysis of \cite{Grijalva}, the configuration of controllable DERs is fixed, and the inverter control laws do not consider newer requirements for having both real and reactive power actuation capability \cite{IEEE1547}. %
Yet the configuration and control law of controllable DERs has a significant impact on whether voltage violations occur \cite{Nazar_placement} and should be considered. DER placement tools should consider those factors and also preserve the visual features of capacity maps to gain topological insights.
Our tool may also be compared to optimal power flow simulations. The approaches in \cite{Mather_monte,farivar_OPF} compute optimal power dispatches for controllable DERs to achieve objectives, but require detailed setup and can be computationally expensive to solve. Furthermore, modeling assumptions made in the choice of lines, loads, and weather may limit the applicability of the simulation results. Rather than setup a detailed simulation, our tool quickly evaluates many controller configurations and only requires a feeder line impedance model. Because our tool does not simulate any system, we do not make claims about time-series performance, DER capacity limit violations, or power loss. Our tool focuses in investigating the stability of proposed control configurations, whose results can then inform the setup of a simulation if needed.

In this work we develop a novel linear state space model to analyze the network-wide stability of inverters that inject real and reactive power. To do so, we assume the dynamics of loads, lines, and inverters are much faster than the inverter power set-point control law \cite{Eggli,Helou,Farivar}. The algebraic DistFlow equations and the proposed integrator controllers form a closed-loop quasi-steady state dynamical system. The proposed system is consistent with the model in \cite{Helou}, with an extension that incorporates the voltage phase angle. 

\section{Problem Formulation}
\subsection{Example on IEEE 123-node Test Feeder}\label{motex_section}
The Phasor-Based Control (PBC) framework involves sending real and reactive power set-point commands to controllable DERs at \emph{actuator nodes} to drive the voltage magnitude and phase angle measured at designated \emph{performance nodes} to computed optimal phasor targets. By achieving these targets, we meet supervisory-level objectives such as reducing voltage volatility, balancing the three phases, and preventing reverse power flow.

We motivate this work by way of an example on the 123NF
\cite{IEEEtest_feed}: 
let $\chi$ be an arbitrarily chosen controller configuration 
of co-located actuator and performance node pairs where PBC successfully drives convergence to the phasor targets.
We create configuration $C_1$ by adding an actuator-performance node pair at node 61 to $\chi$. In contrast, we create configuration $C_2$ by adding a pair to $\chi$ at node 60, which is adjacent to node 61. We design PBC controllers with $C_1$ and $C_2$ using the method from our previous work \cite{GA_paper}, and simulate the closed-loop systems using Opal-RT's ePHASORSIM nonlinear power flow simulator. In the simulation we use reasonable phasor targets and initial conditions, and apply three step change power disturbances. %

\begin{figure}[!h]  
       \centering 
       \includegraphics[width=.4\textwidth]{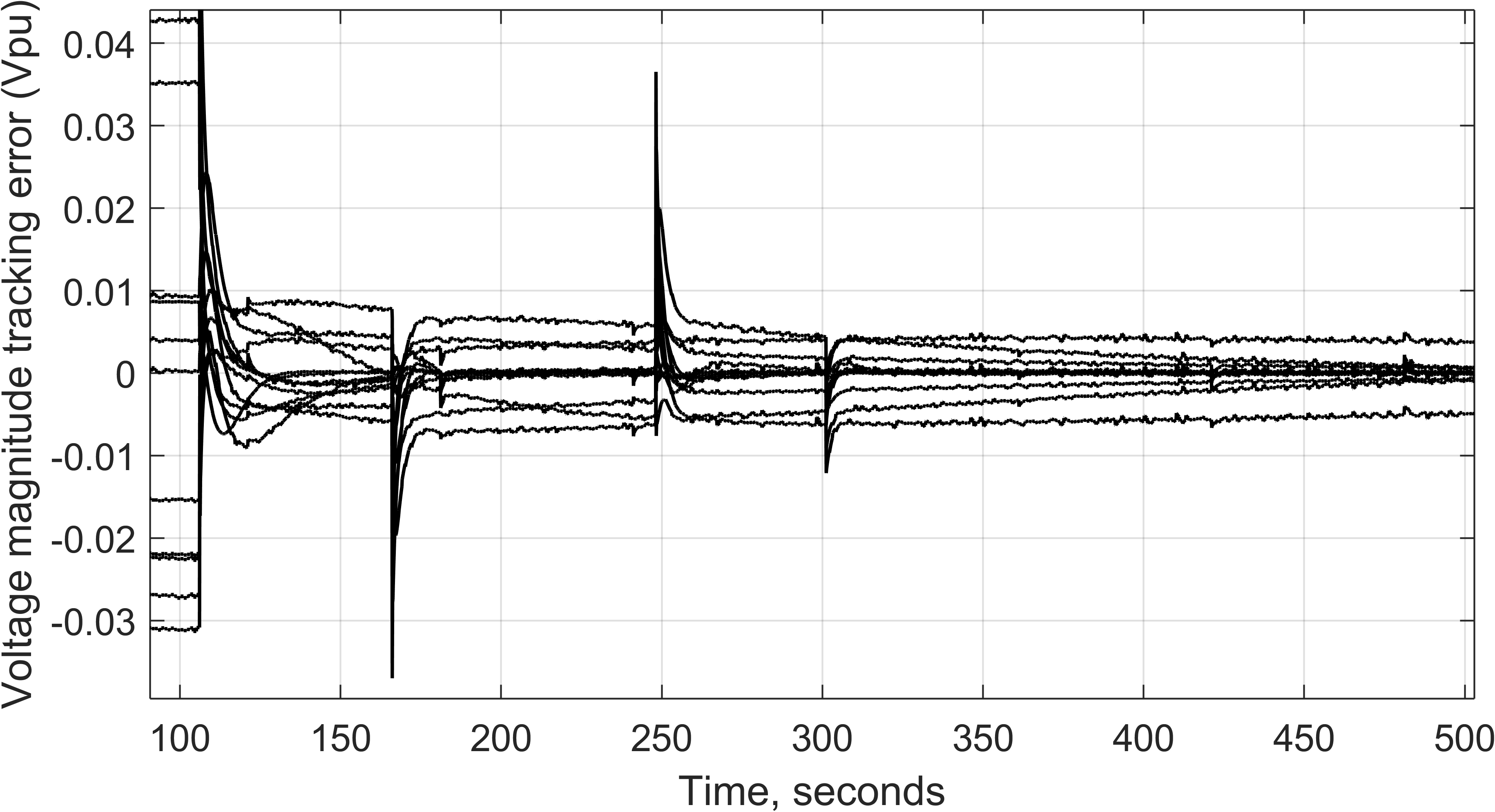}
        \caption{Voltage magnitude tracking error failing to converge when configuration $C_2$ is simulated on the 123NF} \label{motex_60}
\end{figure}

\begin{figure}[!h]  
      \centering 
      \includegraphics[width=.4\textwidth]{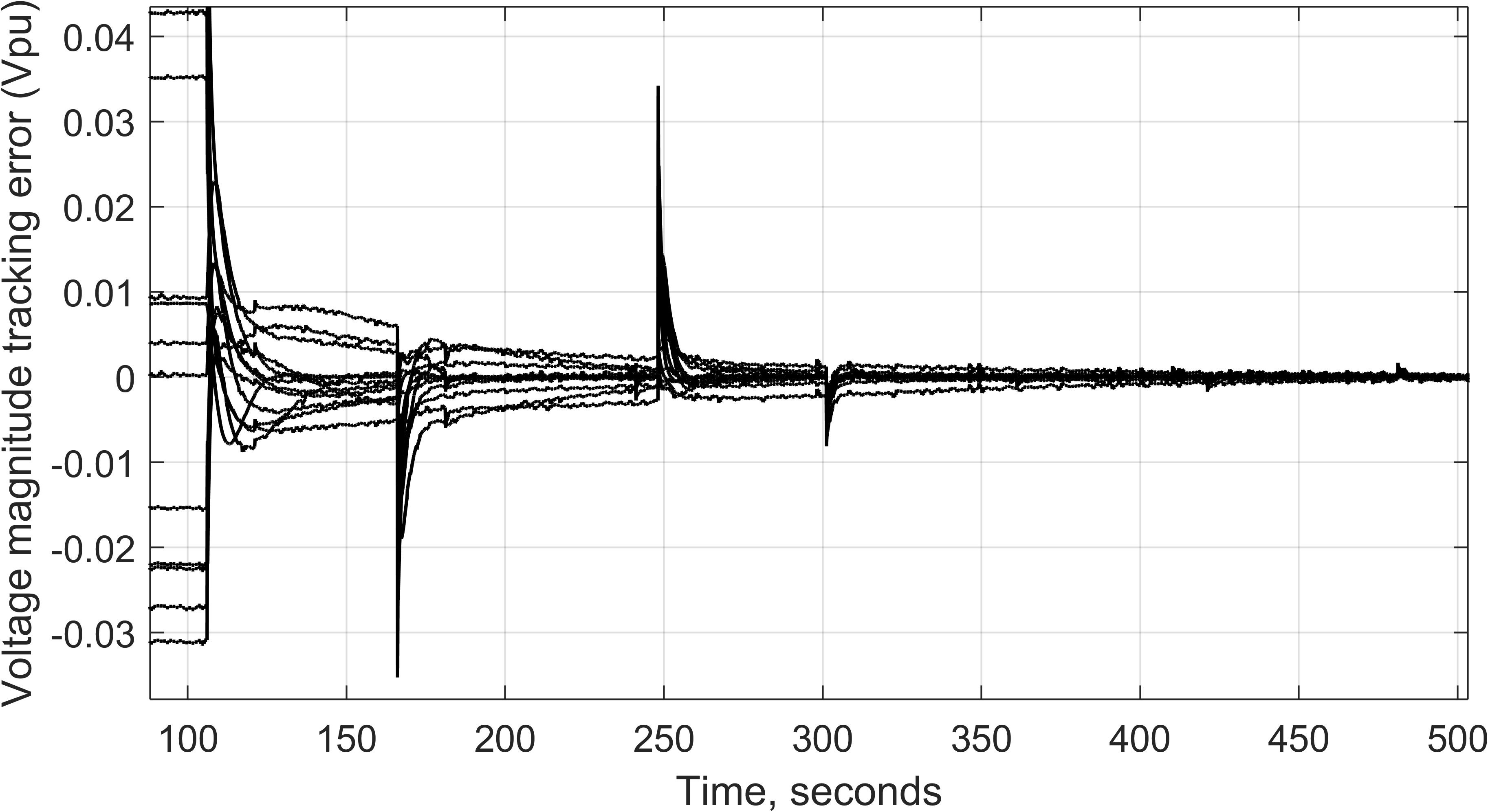}
        \caption{Voltage magnitude tracking error converging when configuration $C_1$ is simulated on the 123NF} \label{motex_61}
\end{figure}

As shown in Fig. \ref{motex_60}, the voltage magnitude tracking errors of $C_2$ do not converge to zero. In contrast, the voltage magnitude tracking errors of $C_1$ converge after 350 seconds to $\leq$0.2\% (five volts), as shown in Fig. \ref{motex_61}. Before the simulation, it was not intuitive whether either configuration would converge. Furthermore, it is not obvious why $C_1$ converges but $C_2$ does not. Our tool can determine, without simulating the scenario, whether a set of controller gains exists that will ensure tracking convergence for any given controller configuration. In our results section we return to this example and employ our tool to provide insight on the differing convergence behavior.

\vspace{-0.2in}

\subsection{State Space Model Derivation}
\label{stateSpaceEqn_section}

\subsubsection{Voltage Magnitude Equation} \label{vmag_sec}
Consider the Distflow \cite{baran1989} branch equation for a single-phase radial network  
\begin{equation}
\label{DistExactMag}
\left|V_i\right|^2 - \left|V_j\right|^2 = 2 (r_{ij}P_{ij} + x_{ij}Q_{ij}) + (r_{ij}^2 + x_{ij}^2) \frac{(P_{ij}^2 + Q_{ij}^2)}{\left|V_i\right|^2},
\end{equation}
which approximates the relationship between voltage magnitudes $V_i, V_j$ and power flow $P_{ij}+\mathbf{j}Q_{ij}$ from node $i$ to node $j$ with complex impedance $r_{ij} + \mathbf{j}x_{ij}$.

We linearize \eqref{DistExactMag} about a nominal voltage of $1 p.u.$ by dropping the square term. Next, let $v_i$ be the squared voltage magnitude, $p_i$ the net real power, $q_i$ the net reactive power at node $i$, and define vectors $v=[v_1,v_2,...v_{n}]^T$, $p=[p_1,p_2,...p_{n}]^T$,
$q=[q_1,q_2,...q_{n}]^T$,$v_0=[v_0,v_0,...v_0]^T$ on a network with $n$ nodes. Here $v_0$ refers to the substation node which is constant at $1 p.u.$. 
 As done in \cite{Eggli,Helou,Farivar}, we choose time step $k$ to be sufficiently large enough for the dynamics from inverters, lines, and loads to settle to steady state between power injection updates. 
The algebraic relationship between nodal power injections and squared nodal voltages at all nodes for time steps $k$ and $k+1$ becomes
\begin{align}
    v_k=Rp_k+Xq_k+v_0 \label{vHelou_k}\\
    v_{k+1}=Rp_{k+1}+Xq_{k+1}+v_0 \label{vHelou_k+1}
\end{align}
where the entries of matrices R and X at the $i^{th}$ row and $j^{th}$ column are given by
\begin{align}\label{RXeqn1}
    R_{ij}=2 \sum_{(h,k)\in \mathcal{P}_i \cap \mathcal{P}_j}^{} r_{hk}\\\label{RXeqn2}
    X_{ij}=2 \sum_{(h,k)\in \mathcal{P}_i \cap \mathcal{P}_j}^{} x_{hk}.
\end{align}

 $\mathcal{P}_i$ is the unique set of lines (or path) connecting node $i$ back to the substation node. 

We subtract equation \eqref{vHelou_k} from \eqref{vHelou_k+1}, giving
\begin{align} 
    v_{k+1}=v_{k}+R(p_{k+1}-p_k)+X(q_{k+1}-q_k).\label{v_written_out1}
\end{align}

To extend \eqref{v_written_out1} to a three-phase system we consider each phase as a separate node and triple the set of $n$ nodes as done in \cite[Appendix]{Eggli}. Each vector element in \eqref{v_written_out1} is replaced with a 3x1 vector, and each element of matrices R and X is replaced with a 3x3 block matrix. This gives $v,p,q \in \mathcal{R}^{3n \times 1}$ and $R,X \in \mathcal{R}^{3n \times 3n}$.

Because we typically seek to drive voltage magnitudes to $1p.u.$ and balance phases to $120 \degree$ apart, we define time-varying phasor targets with $v^{ref} \in \mathcal{R}^{3n}$ values near one, and $\delta^{ref}  \in \mathcal{R}^{3n}$ values with a repeated sequence near $[0 ~-120 ~120]^\top$. Subtracting $v^{ref}_{k+1}$ from both sides of \eqref{v_written_out1}, we have
\begin{multline}
    v_{k+1}-v_{k+1}^{ref}=(v_{k}-v_{k}^{ref})+(v_{k}^{ref}-v_{k+1}^{ref})+\\
    R(p_{k+1}-p_k)+X(q_{k+1}-q_k) \label{v_written_out2}
\end{multline}

\subsubsection{Voltage Phase Angle Equation}
The relationship between nodal voltage phase angles $\delta_i, \delta_j$ and power flow on a single-phase radial network is described by
\begin{equation}
\label{DistExactAngle}
\sin(\delta_i - \delta_j) = \frac{x_{ij}P_{ij} - r_{ij}Q_{ij}}{\left|V_i\right|\left|V_j\right|}.
\end{equation}
To linearize the equation, we make the small-angle approximation $\sin \delta \approx \delta$ and assume $V_i\approx V_j\approx 1 p.u.$, retaining the dependence of $P_{ij}$ and $Q_{ij}$ on $(\delta_i - \delta_j)$ \cite{pbc_journal}.

In extension to all nodes, we define the vector $\delta=[\delta_1,\delta_2,...\delta_{n}]^T$.  Here $\delta_0$ is a vector of the substation nodal voltage phase angle, which is constant at $0\degree$. 
The algebraic relationship between nodal power injections and voltage phase angles at all nodes for time steps $k$ and $k+1$ becomes
\begin{align}
    \delta_{k}=-\frac{1}{2}Rq_k+\frac{1}{2}Xp_k+\delta_0 \label{del_k}\\
    \delta_{k+1}=-\frac{1}{2}Rq_{k+1}+\frac{1}{2}Xp_{k+1}+\delta_0. \label{del_k+1}
\end{align}
We subtract equation \eqref{del_k} from \eqref{del_k+1}, giving
\begin{align}
    \delta_{k+1}=\delta_{k}-\frac{1}{2}R(q_{k+1}-q_k)+\frac{1}{2}X(p_{k+1}-p_k).\label{del_written_out1}
\end{align}
We extend \eqref{del_written_out1} to a three-phase system in the same way as done in section \ref{vmag_sec}.
Then subtracting the phasor targets $\delta^{ref}_{k+1}$ from both sides gives
\begin{multline}
    \delta_{k+1}-\delta_{k+1}^{ref}=(\delta_{k}-\delta_{k}^{ref})+(\delta_{k}^{ref}-\delta_{k+1}^{ref})-\\
    \frac{1}{2}R(q_{k+1}-q_k)+\frac{1}{2}X(p_{k+1}-p_k).\label{del_written_out2}
\end{multline}

\subsubsection{Nodal Power Update Equations}

The update equation for net nodal powers at time step $k$ 
\begin{align}
    q_k=q^{inv}_k+q^{other}_k\\
    p_k=p^{inv}_k+p^{other}_k,
\end{align}
where ($q^{inv},p^{inv}$) are inverter set-point power commands and ($q^{other}, p^{other}$) are power injections from load and generation sources that are not under our control.
We consider these power update equations at time step $k+1$, then substitute these four equations into \eqref{v_written_out2} and \eqref{del_written_out2}, giving
\begin{multline}
    v_{k+1}-v_{k+1}^{ref}=(v_{k}-v_{k}^{ref})+(v_{k}^{ref}-v_{k+1}^{ref})+\\
    R(p^{inv}_{k+1}-p^{inv}_k)+R(p^{other}_{k+1}-p^{other}_k)+\\
    X(q^{inv}_{k+1}-q^{inv}_k)+X(q^{other}_{k+1}-q^{other}_k), \label{stateSpace_long1}
\end{multline}
\begin{multline}
    \delta_{k+1}-\delta_{k+1}^{ref}=(\delta_{k}-\delta_{k}^{ref})+(\delta_{k}^{ref}-\delta_{k+1}^{ref})-\\
    \frac{1}{2}R(q^{inv}_{k+1}-q^{inv}_k)-\frac{1}{2}R(q^{other}_{k+1}-q^{other}_k)+\\
    \frac{1}{2}X(p^{inv}_{k+1}-p^{inv}_k)+
    \frac{1}{2}X(p^{other}_{k+1}-p^{other}_k).  \label{stateSpace_long2}
\end{multline}

\subsubsection{Inverter Control Law}
We design our controller set-point update strategy as a discrete-time integrator with gain matrices $F_{11}, F_{12}, F_{21},$ and $F_{22}$, where 
\begin{align}
    q^{inv}_{k+1}-q^{inv}_k=-F_{11} (v_k-v^{ref}_k) -F_{12}(\delta_k-\delta^{ref}_k),\label{ulaw1}\\
    p^{inv}_{k+1}-p^{inv}_k=-F^{21}(v_k-v^{ref}_k) -F_{22}(\delta_k-\delta^{ref}_k). \label{ulaw2} %
\end{align}
The inverter power injection will decrease when the voltage is too high $(v_k-v_k^{ref})>0$ and increase when voltage is too low. Equation \eqref{ulaw1} differs from the analogous DVVC law, $q^{inv}_{k+1}=-F_{11}(v_k-v^{ref}_k)$. Specifically, we map the tracking error to the next \emph{change in actuation} command, in contrast to how DVVC maps it to the next actuation command. 

\subsubsection{Quasi-steady State Dynamical System}
Let our states be the voltage magnitude tracking error $e^v=v-v^{ref}$ and voltage phase angle tracking error $e^\delta=\delta-\delta^{ref}$. We define our inputs be the change in inverter actuation $u^q_k=q^{inv}_{k+1}-q^{inv}_k$, $u^p_k=p^{inv}_{k+1}-p^{inv}_k$. After substituting \eqref{ulaw1} and \eqref{ulaw2} into \eqref{stateSpace_long1} and \eqref{stateSpace_long2}, our state space equations are

\begin{gather} 
 \begin{bmatrix} e^v_{k+1} \\ e^\delta_{k+1} \end{bmatrix}
 =A
   \begin{bmatrix} e^v_{k} \\ e^\delta_{k} \end{bmatrix}   +
     B
        \begin{bmatrix}   u_k^q \\ u_k^p  \end{bmatrix}   +
     \begin{bmatrix}   c_k^q \\ c_k^p  \end{bmatrix}   +
     \begin{bmatrix}   d_k^q \\ d_k^p  \end{bmatrix}, \label{OLblock}
\end{gather} 
\begin{gather} 
 \begin{bmatrix} u_k^q \\ u_k^p    \end{bmatrix}
    =-F
  \begin{bmatrix} e^v_{k} \\ e^\delta_{k}\end{bmatrix}, \label{stateFbblock}
  \end{gather}
  \begin{gather}
   A =
  \begin{bmatrix}
   I & 0 \\
    0 & I\\
   \end{bmatrix}, B=
     \begin{bmatrix}
   X & R \\
    -\frac{1}{2}R & \frac{1}{2}X\\
   \end{bmatrix}, F=
\begin{bmatrix}
    F_{11} & F_{12} \\
    F_{21} & F_{22} \\
   \end{bmatrix},
\end{gather} 

where $c_k^v=(v_{k}^{ref}-v_{k+1}^{ref})$, and $d^q_k=R(p^{other}_{k+1}-p^{other}_k) +X(q^{other}_{k+1}-q^{other}_k)$, which are the changes in voltage magnitude targets and changes in voltage magnitude from uncontrollable sources, respectively. Likewise, $c_k^\delta=(\delta_{k}^{ref}-\delta_{k+1}^{ref})$, and $d^p_k=\frac{1}{2}X(p^{other}_{k+1}-p^{other}_k) -\frac{1}{2}R(q^{other}_{k+1}-q^{other}_k)$. Note that $c^q_k,c^p_k,d^q_k$, and $d^p_k$ are time-varying terms that are independent of the state. The authors in \cite{Helou} formulate a similar open loop and feedback control equations, but only for reactive power driving voltage magnitudes to the nominal $V=1 p.u.$ on single-phase networks. Our use of power injections to track voltage phase angle targets is novel as well as critical to the PBC framework.

\subsection{Disturbance Model}

After substituting the feedback law \eqref{stateFbblock} into \eqref{OLblock}, we have $x_{k+1}=(A-BF)x_k+c_k+d_k$.

Power disturbances from uncontrollable sources such as load changes, cloud cover events, and solar PV fluctuations can cause voltage spikes resulting in unintended device tripping. We model these disturbances with time-series profiles as done in \cite{Helou,Eggli}. Modeled this way, disturbances $d_k$ and phasor targets $c_k$ do not change the eigenvalues of $A_{cl}$. To evaluate stability of the closed-loop system, we set $c_k=d_k=0$, giving
\begin{align}\label{CLsys}
    x_{k+1}=(A-BF)x_k=A_{cl}x_k.
\end{align}
 Equation \eqref{CLsys} is a linear time-invariant (LTI) system with dynamic elements arising from integral control action.

In what follows we design $F$ so that the system $x_{k+1}=A_{cl}x_k+c_k+d_k$ operates close to the equilibrium of $V=1 pu$. In more detail, we assume the phasor target changes $c_k$, voltage disturbances $d_k$, and initial conditions are small, such that the open-loop system \eqref{OLblock} operates close to our linearization equilibrium of $V=1 pu$.

\subsection{How Configurations Inform Gain Matrix Requirements}
Communication setups of controllable DERs and sensors on real grids directly determine sparsity requirements on the controller gain matrix $F$. For a three-phase grid network with $n$ nodes, $F$ has dimension $6n \times 6n$. 

In $F_{11}, F_{12}, F_{21},$ and $F_{22}$, there is an arrangement of $n^2$ 3x3 blocks, each one representing a node with up to three phases. If an actuator on node $i$ is actuating to track the phasor target at node $j$, there is a nonzero 3x3 block at the (i,j) location of $F_{11}, F_{12}, F_{21},$ and $F_{22}$.

If in our control law, reactive power commands are computed as a function of voltage magnitude and not phase angle, $F_{12}=0$. Similarly, if real power commands are computed as a function of phase angle and not voltage magnitude, $F_{21}=0$.
If power commands on one phase are computed from only that phase's measurements, then all 3x3 blocks are diagonal.

Considering realistic scenarios of the PBC framework, we make the following assumptions when designing $F$.
\begin{assumption} \label{assump_Frows}
    each actuator is used to track a single phasor target. As a result, every row of $F$ can only have up to one non-zero element.
\end{assumption}
\begin{assumption} \label{assump_SISO}
    $F_{12}=F_{21}=0$, and the 3x3 blocks representing three-phase nodes in $F_{11}$ and $F_{22}$ are diagonal.
\end{assumption}

 Let the subspace $W$ encapsulate all sparsity structure requirements on matrix F. Let $ I_W$ be the structural identity of the subspace S, with its $ij^{th}$ entry defined as
\begin{align}
    [I_W]_{ij} =  \begin{cases} 
      1, & \text{if} \quad F_{ij} \quad \text{is a free variable;} \\
      0, & \text{if} \quad F_{ij}=0 \quad \text{is required.} 
   \end{cases}
\end{align}
We seek an $F \in W$ that drives \eqref{CLsys} to be closed-loop stable.

\subsection{Applicability of Controllability}

One may seek to assess controllability of $(A,B)$ from \eqref{OLblock}.
If $(A,B)$ is controllable, there exists an $F$ matrix of \emph{any} sparsity structure that makes \eqref{CLsys} stable. $F$ could be totally dense, which corresponds to an unrealistic communication setup where there is a controllable actuator at \emph{every} node, and each one has access to sensor measurements at \emph{every} node. Realistic communication setups are likely to impose sparsity requirements, such as those in Assumption \ref{assump_Frows}. For example, with DVVC inverters, where inverters actuate reactive power to regulate the voltage magnitude their own nodes, the $F$ matrix would require $F_{11}$ to be diagonal and $F_{12}=F_{21}=F_{22}=0$. Therefore, controllability of \eqref{OLblock} is not generally sufficient for determining the convergence of realistic configurations of actuators.

\subsection{Conditions for Stability}
We use $|.|$ as the cardinality operator on sets. Let $M(i,:)$ denote the $i^{th}$ row, and $M(:,i)$ the $i^{th}$ column of a matrix M . If $i$ is a set of indices then $M(i,:)$ and $M(:,i)$ represent the rows and columns at those indices, respectively. %

A discrete LTI system $x_{k+1}=M x_k$ is said to be \emph{stable in the sense of Lyapunov} (SISL) if $x_k$ is bounded for all time $k\geq k_0$. If the eigenvalues of $M$ are on or in the unit circle, and those on the unit circle have 1x1 Jordan blocks, the system is SISL \cite[Chapter 5.3]{Chen_linsys}. 
It is computationally expensive to compute the Jordan canonical form of a matrix. Instead, we leverage the property that the number of Jordan blocks associated with an eigenvalue $\lambda$ of $M$ is equal to the nullity of $(M-\lambda I)$. If the multiplicity of $\lambda$ is equal to this nullity, all associated Jordan blocks are 1x1. Let U be each distinct set of eigenvalues on the unit circle. 

\begin{theorem}If
    \begin{enumerate}
        \item $abs(\lambda)\leq 1 \quad \forall \lambda \in eig(M)$
        \item nullity$(M-\lambda I)=|U|\quad \forall \lambda \in U$
    \end{enumerate}
    then the system $x_{k+1}=M x_k$ is stable in the sense of Lyapunov. \label{stability_thm}
\end{theorem}

\subsection{Stability Yields Phasor Tracking}

 We now prove that if \eqref{CLsys} satisfies Theorem \ref{stability_thm}, the inverters will drive the performance node voltages to their phasor targets. Consider the phasor tracking error $x \in \mathcal{R}^{3n}$ from \eqref{CLsys}. Let $p$ be the indices of $x$ that we want to drive to zero. Let $\bar{p}$ be the remaining indices of $x$. Note that $|p|+|\bar{p}| = n$. 

\begin{proposition}\label{stab_claim}
Consider a gain matrix $F$ defined in \eqref{stateFbblock} whose columns at indices p are not the zero vector ($F(:,p)\neq0$). If $F$ is designed so that \eqref{CLsys} satisfies Theorem \ref{stability_thm}, the ${p}$-indexed states $x(p)$ will be driven zero.
\end{proposition}

The state vector of \eqref{CLsys} can be written with Jordan canonical form as
\begin{align}
    x_k&=A_{cl}^k x_0=TJ^kT^{-1}x_0.
\end{align}
The right eigenvectors of $A_{cl}$ , $e_i$, make up the columns of $T$, the transpose of the left eigenvectors $v_i$ make up the rows of $T^{-1}$, and $J$ is the Jordan block matrix of $A_{cl}$ \cite{CalandDes}.
Let $U$ denote the set of eigenvalues at one, and let $S$ denote those inside the unit circle. Because the open-loop system in \eqref{OLblock} has all eigenvalues equal to one, a SISL closed-loop system in \eqref{CLsys} will have eigenvalues in $U$ or in $S$.

 The expansion of $x_k$ into eigenvalue terms is
\begin{align}
    x_k&=\sum_{i \in S}^{}{\tilde{E_i} \tilde{\Pi}_i^k \tilde{V_i}^Tx_0} + \sum_{j \in U}^{}{E_j \Pi_j^k V_j^Tx_0},\label{dyadic}\\ 
    &=\sum_{i \in S}^{}{s_{i,k}} + \sum_{j \in U}^{}{u_{j,k}}, \label{su_terms}
\end{align}
where $E_j$ ($\tilde{E_i}$) is constructed by selecting the submatrix of $T$ associated with the $j^{th}$ ($i^{th}$) eigenvalue, then zero-padding the submatrix to obtain the same dimensions as $T$. In the same way, $V_j$ ($\tilde{V_i}$) constructed from $T^{-1}$, and $\Pi_j^k$ ($\tilde{\Pi}_i^k$) is constructed from $J^k$. Details of this construction can be found in \cite[Chapter 4.4]{CalandDes}. We denote $u_{j,k}$ ($s_{i,k}$) as the $j^{th}$ ($i^{th}$) term associated with eigenvalues equal to one (in the unit circle).

\begin{lemma}\label{marg_stab_lemma}
    If the closed-loop system \eqref{CLsys} is designed to satisfy Theorem \ref{stability_thm}, $F(:,p)\neq0 \implies E_j(p,:)=0 \quad\forall j \in U$.
\end{lemma}
  The proof of Lemma \ref{marg_stab_lemma} is in the Appendix.

Proof of Proposition \ref{stab_claim}: Over time $\tilde{\Pi}_i^k \rightarrow 0$, causing $s_{i,k} \rightarrow 0$. In general the marginally stable eigenvalue terms $u_{j,k}$ contribute a constant value to all states, preventing any state from approaching zero. However, by Lemma \ref{marg_stab_lemma}, $E_j(p,:)=0 \quad \forall j \in U$. Therefore $u_{j,k}(p)=0 \quad \forall k \geq 0$. Together with $s_i^k \rightarrow 0$, $x_k(p) \rightarrow 0$.

\subsection{Existence of Viable Controllers}

\subsubsection{The Existence Problem} 
One can determine whether there exists an unconstrained $F$ matrix that satisfies the Lyapunov equation by solving a convex semi-definite program (SDP) \cite[Chapter 4.4]{solving_LMIs}. However, determining whether there exists an $F \in W$ that satisfies the lyapunov equation is a harder problem, and is sometimes setup as the static output feedback (SOF) stabilization problem \cite{SOF3,SOF1}. This problem is a nonlinear (bilinear) SDP problem that is NP-hard \cite{SOF1}. Because few nonlinear SDP solvers produce reliable results for scaling system sizes, we instead sample candidate $F$ matrices in a parameter space for our tool.

\subsubsection{Sampling $F$ matrices} \label{sampling_sec}

For a given configuration if we can find at least one $F \in W$ such that $x_{k+1}=A_{cl}x_k$ satisfies Theorem \ref{stability_thm}, we call the configuration and associated $F$ matrix \emph{good}. Searching for good $F$'s can never be exhaustive, so a \emph{poor} configuration is one where we did not find a suitable $F$.

We now define a parameter space to meet design Assumptions \ref{assump_Frows} and \ref{assump_SISO}. Within each of $F_{11}$, $F_{12}$, $F_{21}$, and $F_{22}$ we choose all non-zero gain elements to be equal positive values $f$. From experience, when all gain elements are non-negative, $F$ matrices with positive elements near the origin tend to be good. Hence we define the parameter space as a positive cone around the origin:
\begin{align} \label{Fsamp}
    [F_{11}]_{ij},[F_{21}]_{ij}=  \begin{cases} 
      0, & \text{if} \quad [I_W]_{ij}=0 \\
      f_q \in [0,F^q_{ub}], & \quad otherwise,
      \end{cases}
\end{align}
\begin{align}
    [F_{12}]_{ij},[F_{22}]_{ij}=  \begin{cases} 
      0, & \text{if} \quad [I_W]_{ij}=0 \\
      f_p \in [0,F^p_{ub}], & \quad otherwise,
      \end{cases}
\end{align}
where $F^q_{ub}=(1/q)$ and $F^p_{ub}=(1/p)$ is computed once for each configuration. 
We compute $q$ ($p$) as a heuristic estimate of the ratio of reactive power (real power) change at the actuation node to voltage magnitude (voltage phase angle) change at the associated performance node, averaged across all actuator-performance node pairs in the configuration.

\section{Results}
 The procedure in Section \ref{sampling_sec} allows us to determine the stability of any configuration of actuator-performance node pairs (APNPs). We now present three processes of our tool that exercise this stability assessment on the 123NF, noting that the tool has also been implemented on a 344-node feeder. For all configurations we adopt Assumptions \ref{assump_Frows} and \ref{assump_SISO}, then change the locations of APNPs to enforce different structural requirements on $F$ per configuration. An APNP is \emph{co-located} if an actuator and performance node are at the same location, and \emph{non-colocated} otherwise. 
 
 Figures and code associated the following results can be downloaded at: \url{https://github.com/jaimiosyncrasy/heatMap}.
 
\subsection{Non-colocated Placement Process (NPP)}
The NPP is useful when users cannot place controllable DERs at the same locations as problematic voltages. Additionally, it can be valuable to coordinate multiple actuators at different nodes to track a voltage at another node.

The NPP can be applied to a grid with any number of existing APNPs. We choose a candidate performance node location, then iterate through every other empty node in the feeder, fixing each as the associated candidate actuator node. We generate a heatmap of the network, where a node's color indicates the stability of the configuration created by appending the candidate APNP to the existing set of APNPs. If the node is blue, several (at least 7\% of) $F$ matrices sampled from the parameter space defined in Section \ref{sampling_sec} were found that make the closed-loop system \eqref{CLsys} satisfy Theorem \ref{stability_thm}. If the node is yellow, only a few (less than 7\%) were found, and if red, no $F$ matrices were found. Next we select a blue or yellow APNP to become part of the core configuration, and this chosen APNP is colored grey on subsequent heatmaps. We repeat this process until the desired number of APNPs have been placed. The NPP produces a stable configuration of APNPs in desired locations, and setp-by-step heatmaps show how the choice of performance node and placed APNPs affect the stability of placing the next APNP.

\begin{figure}[!h]  
       \centering 
       \includegraphics[width=.40\textwidth]{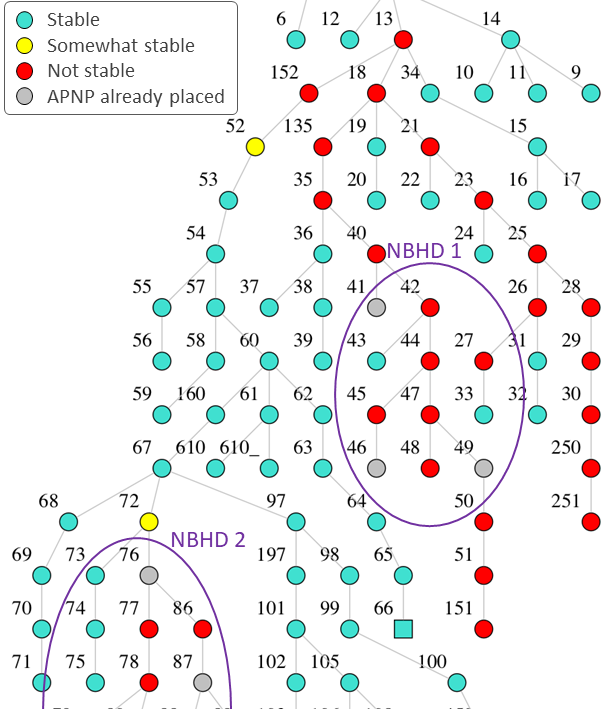}
        \caption{Heatmap generated by the NPP indicating good locations for placing a seventh actuator for tracking the voltage phasor at node 66 on the 123NF.}
        \label{nbhd_colormap}
\end{figure}

\subsubsection{Neighborhood Scenario}
We demonstrate the NPP on a feeder where there are two neighborhoods of existing APNPs. Specifically, a group of actuators located at node 41, 46, and 49 track the voltage phasor at node 44, and a group of actuators at node 76, 82, and 87 track the voltage phasor at node 77. In total there is an existing configuration of six APNPs. We run the NPP to find good locations for placing a seventh APNP, which will start a new group of actuators for tracking the voltage phasor at the chosen performance node (node 66). As shown in the heatmap of Fig. \ref{nbhd_colormap}, locations near node 66 (marked with square), while locations near the existing two neighborhoods of actuators are more yellow and red. Notably, there is a color gradient between performance node 66 and performance node 44, but predicting the location of this color transition without the heatmap is not intuitive. 

We validate the heatmap colors at nodes 152, 54, 67, 72, 84, and 18 from Fig. \ref{nbhd_colormap} using ePHASORsim. For blue and yellow nodes, we use the associated stable $F$ matrices to simulate the closed-loop system \eqref{CLsys} on the nonlinear grid model, and verify that the system converges to the phasor target. For red nodes, we use a more advanced but slower parameter search algorithm from \cite{GA_paper} than our method in Section \ref{sampling_sec}. All but one of the nodes checked by ePHASORSIM agree with the NPP results. Node 152 should be yellow and not red, indicating that our tool errs on the conservative side for borderline yellow-red cases.

\subsubsection{Branch Analysis}

\begin{table}[]
\caption{\label{tab:branch_table}Best branches on the 123NF from running the NPP on the Neighborhood Scenario.}
\begin{center}
\begin{tabular}{|p{2.5cm}|p{1.2cm}|p{1.7cm}|}
\hline
\multicolumn{1}{|c|}{Branch start and end node} & Percent stable & branch length (no. nodes)\\ \hline
node\_68-node\_71               & 100 & 4                        \\ \hline
node\_109-node\_114             & 100 & 5                        \\ \hline
node\_197-node\_350               & 100 & 6                        \\ \hline
node\_152-node\_66               & 98.7 & 11                        \\ \hline
node\_160-node\_451              & 98.2 & 8                        \\ \hline
\end{tabular}
\end{center}
\end{table}

Data collected across steps of the NPP can indicate good feeder branches to place APNPs. Let $S$ be the set of 348 stable configurations found by the NPP when placing one to seven APNPs in the above neighborhood scenario. We call a network branch \emph{used} each time at least one actuator from a stable configuration in $S$ is located on that branch. Then for each branch, we compute the percentage of times it is used out of the total number of configurations involving the branch. This metric captures the branches that are the \emph{most blue} across the NPP heatmaps. Table \ref{tab:branch_table} lists the five branches with the highest percentage metric among the branches with length of at least four nodes. The best branches are located in the region between neighborhood groups one and two.
\subsection{Overload Co-located Placement Process (OCPP)}
 We randomly choose a node with no APNP, and if adding a co-located APNP there maintains a stable configuration, we place it there. We repeat this until our random choice yields an unstable configuration. We generate a heatmap with the same color key as the NPP that shows the stability of the configuration that would result from adding a co-located APNP at each of the remaining feeder nodes.

Running the OCPP with a random seed of 3 generates the same configuration as $\chi$ from section \ref{motex_section}. 
 We define the feeder's \emph{main branch} to be the path from the edge node to the substation that has the most nodes branching from the path.  The main branch of the 123NF is node\_1-node\_96. We observe from the OCPP heatmap that this branch has many red and yellow nodes compared to other branches, indicating that it is better to place co-located actuators away from the main branch. Notably, node 60 is on the main branch, while node 61 is not, which may indicate why $C_1$ converged but $C_2$ did not in section \ref{motex_section}.
\subsection{Auto Overload CPP (Auto-OCPP)}
We randomly choose a node with no APNP, and if adding a co-located APNP there maintains a stable configuration, we place it there. In contrast to the non-auto OCPP, when our random choice of location yields an unstable configuration, instead of stopping and producing a heatmap we continue to randomly sample the remaining nodes until we find a stable APNP. This algorithm finished when there are no remaining nodes that would maintain a stable configuration. 

\begin{table}[]
\caption{\label{tab:camel} Last four APNPs before all nodes unstable from six trials of running the Auto-OCPP.}
\begin{tabular}{|p{1cm}|p{2cm}|p{4cm}|}
\hline
\multicolumn{1}{|c|}{seed} & total placed & nodal distance to substation for last four APNPs (fourth-last, third-last, second-last, last) \\ \hline
6                          & 11               & 1, 4, 13, 12                        \\ \hline
5                          & 12               & 16, 10, 11, 4                        \\ \hline
2                          & 13               & 16, 10, 3, 2                   \\ \hline
8                          & 25               & 9, 17, 19, 4                        \\ \hline
4                          & 27               & 16, 12, 2, 5                       \\ \hline
3                          & 34               & 14, 5, 2, 2                     \\ \hline
\end{tabular}
\end{table}

\begin{table}[]
\caption{\label{tab:overload} Topology information about APNPs placed, from six trials of running the Auto-OCPP.}
\begin{tabular}
{|p{0.5cm}|p{0.8cm}|p{1.3cm}|p{0.8cm}|p{0.8cm}|p{1.3cm}|}
\hline
\multicolumn{1}{|c|}{seed} & total placed & number on or one-node-from edge & number at fork & number in middle & Percent on or one-node-from edge \\ \hline
6                          & 11               & 4              & 3                       & 4                & 36.4              \\ \hline
5                          & 12               & 8              & 2                       & 2                & 66.7              \\ \hline
2                          & 13               & 8              & 2                       & 3                & 61.5              \\ \hline
8                          & 25               & 17              & 3                       & 5                & 68.0              \\ \hline
4                          & 27               & 16             & 6                       & 5                & 59.3              \\ \hline
3                          & 34               & 24             & 6                       & 4                & 70.6              \\ \hline
\end{tabular}

\end{table}

We run the Auto-OCPP for six random seed trials, and compare the configurations of each trial in Table \ref{tab:camel}. We observe that the total number of APNPs placed varies significantly. Table \ref{tab:camel} also lists the nodal distance to the substation of the last four co-located APNPs placed before all nodes were unstable. Given that the farthest nodal distance possible is 22 nodes, we observe that the last APNPs are close to the substation, with the exception of the seed 6 trial.

For each run of the Auto-OCPP, we tally the number of APNPs that were placed on or one-node away from an edge node, at a network fork, or in the middle of a branch. From Table \ref{tab:overload}, we observe that the trials with more total APNPs have a higher percentage of APNPs on edge nodes. Placing co-located APNPs at edge nodes may pose the least disruption to the existing configuration, thereby enabling the placement of additional co-located APNPs.

\section{Conclusion}
\label{sec:Conclusion}

In this work we derive a state space model under the PBC framework. We describe why controllability and the Lyapunov equation are not sufficient methods to assess the stability of realistic controller communication setups. We show that if our closed-loop system is designed to be stable in the sense of Lyapunov, the voltages at chosen performance nodes are driven to their phasor targets. We then use the LTI model in a tool to evaluate configurations on the 123NF. For configurations in which inverters regulate voltage phasors at nodes other than their own, we observe that actuators should be spaced apart, and this spacing can be visualized with our tool's heatmaps. Results on the 123NF also indicate that it is good to place co-located actuator-performance node pairs on or near edge nodes, poor to place them on the main branch, and poor to place them near the substation.

In future work we will investigate the modeling of dynamics from lines and loads that have comparable timescales to the inverter set-point control law. We will also compare the design choice of Assumption \ref{assump_SISO} to other state feedback strategies such as volt-watt control. Finally, we will explore conditions on the controller gains to ensure the closed-loop system stays near the linearization equilibrium.

\section*{Acknowledgement}
We thank Alex Devonport for his insightful
discussions. This work is supported by the Department of Energy EERE Office, award number DE-EE0008008. 

\bibliography{library.bib}
\bibliographystyle{ieeetr}

\section{Appendix: Proof of Lemma \ref{marg_stab_lemma}}

Consider matrices $A$, $B$, $F$ for the system in \eqref{CLsys}. The eigenvector equation for the $j^{th}$ one eigenvalue of $A_{cl}$ in $U$ is $(A_{cl}-1 \cdot I)e_j=0$. Because $A=I$, this becomes $(I-BF-I)e_j=0$, which simplifies to $-BFe_j=0$.

Next we show that $B$, which is constructed from $R$ and $X$ matrices in \eqref{OLblock}, is invertible. By definition $X \succ 0$ and $R \succ 0$ \cite{Farivar}.  From properties of positive definite matrices, $\frac{1}{2}X-(-\frac{1}{2}R^T)X^{-1}R \succ 0$. This is useful because $det(B)=det(X)\cdot det(\frac{1}{2}X-(-\frac{1}{2}R^T)X^{-1}R)$.
Thus $det(B)\neq 0 \Leftrightarrow$ B is invertible.

Because B is invertible, $\mathcal{N}(B)$ is trivial, so $-BFe_j=0 \implies Fe_j=0$. Next we construct $\tilde{F}$ and $\tilde{e_j}$ by rearranging the columns and components of $F$ and $e_j$ respectively:
\begin{align}
    \tilde{F}=\begin{bmatrix}
        F(:,p) & F(:,\bar{p})
    \end{bmatrix}, ~\tilde{e_j}=
    \begin{bmatrix}
        e_j(p) & e_j(\bar{p})
    \end{bmatrix}^T
\end{align}
Note $\bar{p}$ are the indices of states that are not tracked, so $F(:,\bar{p})=0$. Thus, $Fe_j=\tilde{F}\tilde{e_j}=F(:,p)\cdot e_j(p)$. 
Assumption \ref{assump_Frows} implies that the p-index columns of $F$ are linearly independent, so $\mathcal{N}(F(:,p))$ is trivial. This gives $F(:,p)\cdot e_j(p)=0 \implies e_j(p)=0$. $E_j$ is constructed from the column vectors $e_j$, so $e_j(p)=0 \implies E_j(p,:)=0$

\end{document}